\documentclass[aps,showpacs,nofootinbib,superscriptaddress]{revtex4}

\usepackage{graphicx}
\usepackage{dcolumn}

\def\slashchar#1{\setbox0=\hbox{$#1$}
   \dimen0=\wd0 \setbox1=\hbox{/} \dimen1=\wd1
   \ifdim\dimen0>\dimen1 \rlap{\hbox to \dimen0{\hfil/\hfil}} #1
   \else  \rlap{\hbox to \dimen1{\hfil$#1$\hfil}} / \fi}
\def\esp{\kern 9mm} 

\begin{document}
\title{Charged and Neutral Current Neutrino Induced Nucleon Emission
Reactions}  %
\author{J. Nieves}
\affiliation{Departamento de F\'{\i}sica At\'omica, Molecular y Nuclear, 
          Universidad de Granada,
          Granada 18071, Spain}
\author{M. Valverde}
\affiliation{Departamento de F\'{\i}sica At\'omica, Molecular y Nuclear, 
          Universidad de Granada,
          Granada 18071, Spain}
\author{M.J. Vicente Vacas}
\affiliation{Departamento de F\'\i sica Te\'orica and IFIC,
Centro Mixto Universidad de Valencia-CSIC\\ Institutos de Investigaci\'on
de Paterna, Aptdo. 22085, E-46071 Valencia, Spain }

\begin{abstract}
  \rule{0ex}{3ex} By means of a Monte Carlo cascade method, to account
for the rescattering of the outgoing nucleon, we study the charged and
neutral current inclusive one nucleon knockout reactions off nuclei
induced by neutrinos. The nucleon emission process studied
here is a clear signal for neutral--current neutrino driven reactions, and
can be used in the analysis of future neutrino experiments.

\end{abstract}
  
\maketitle

\section{Introduction and Theoretical Framework}

Recently an interest in neutrino scattering off nuclei has raised
because of its implications in the experiments on neutrino
oscillations based on large Cerenkov detectors. The presence of
neutrinos, being chargeless particles, can only be inferred by
detecting the secondary particles they create when colliding and
interacting with matter. Nuclei are often used as neutrino detectors,
thus a trustable interpretation of neutrino oscillation data heavily
relies on detailed and quantitative knowledge of the features of the
neutrino-nucleus interaction~\cite{RCNN}. For instance, in the case of
neutrino processes driven by the electroweak Neutral Current (NC), the
energy spectrum and angular distribution of the ejected nucleons are
the unique observables. Besides, the study of these distributions
might also help to improve on our knowledge of the quark and gluon
substructure of the nucleon, and in particular on the amount of
nucleon spin carried by strange quarks~\cite{finese}.

 At intermediate energies, above the nuclear giant resonance and below
 the $\Delta(1232)$ regions, neutrino--nucleus interactions have been
 studied within several approaches. Several different Fermi gas,
 Random Phase Approximation (RPA) and shell model  based
 calculations have been developed during the last 15
 years~\cite{Garvey:1992qp}--\cite{Gr04} to compute neutrino or
 antineutrino induced single--nucleon emission cross sections. Most of
 the calculations use the plane wave and distorted wave impulse
 approximations (PWIA and DWIA, respectively), including or not
 relativistic effects.  The PWIA constitutes a poor approximation,
 since it neglects all types of interactions between the ejected
 nucleon and the residual nuclear system. The DWIA describes the
 ejected nucleon as a solution of the Dirac or Schr\"odinger equation
 with an optical potential obtained by fitting elastic proton--nucleus
 scattering data. The imaginary part accounts for the absorption into
 unobserved channels. This scheme is incorrect to study nucleon
 emission processes where the state of the final nucleus is totally
 unobserved, and thus all final nuclear configurations, either in the
 discrete or on the continuum, contribute.  The distortion of the
 nucleon wave function by a complex optical potential removes all
 events where the nucleons collide with other nucleons. Thus, in DWIA
 calculations, the nucleons that interact are lost when in the
 physical process they simply come off the nucleus with a different
 energy, angle, and maybe charge, and they should definitely be taken
 into account. A clear example which illustrates the deficiencies of
 the DWIA models is the neutron emission process: $(\nu_l, l^-
 n)$. Within the impulse approximation neutrinos only interact via
 Charged Current (CC) interactions with neutrons and would emit
 protons, and therefore the DWIA will predict zero cross sections for
 CC one neutron knock-out reactions. However, the primary protons
 interact strongly with the medium and collide with other nucleons
 which are also ejected. As a consequence there is a reduction of the
 flux of high energy protons but a large number of secondary nucleons,
 many of them neutrons, of lower energies appear.

In this talk, we present results for the QE $(\nu_l,\nu_l N)$,
$(\nu_l,l^- N)$, $({\bar \nu}_l,{\bar \nu}_l N)$ and $({\bar
\nu}_l,l^+ N)$ reactions in nuclei. We use a Monte Carlo (MC)
simulation method to account for the rescattering of the outgoing
nucleon.  The first step is the gauge boson ($W^{\pm}$ and $Z^0$ )
absorption in the nucleus, we take this reaction
probability\footnote{It is given by the inclusive QE cross sections
$d^2\sigma /d\Omega^{\prime} dE^{\prime}$ ($\Omega^{\prime}$,
$E^{\prime}$ are the solid angle and energy of the outgoing lepton,
for a fixed incoming neutrino or antineutrino Laboratory (LAB)
energy. We also compute differential cross sections with respect to
$d^3r$. Thus, we also know the point of the nucleus where the gauge
boson was absorbed, and we can start from there our MC propagation of
the ejected nucleon} from the microscopical many body framework
developed in Refs.~\cite{NAV05,NVV06} for CC and NC induced
reactions. Starting from a Local Fermi Gas (LFG) picture of the
nucleus, which automatically accounts for Pauli blocking, several
nuclear effects are taken into account in that scheme: {\it i)} a
correct energy balance, using the experimental $Q-$values, is
enforced, {\it ii)} Coulomb distortion of the charged leptons is
implemented by using the so called ``modified effective momentum
approximation'', {\it iii)} medium polarization (RPA), including
$\Delta-$hole degrees of freedom and explicit pion and rho exchanges
in the vector--isovector channel of the effective nucleon--nucleon
force, and Short Range Correlation (SRC) effects are computed, and
finally {\it iv)} the nucleon propagators are dressed in the nuclear
medium, which amounts to work with a LFG of interacting nucleons and
it also accounts for reaction mechanisms where the gauge boson,
$W^\pm$ or $Z^0$, is absorbed by two nucleons.  This model is a
natural extension of previous studies on electron~\cite{GNO97},
photon~\cite{CO92} and pion~\cite{OTW82,pion} dynamics in nuclei and
predicts QE neutrino--nucleus (differential and integrated) cross
sections with an accuracy of about 10-15\%, at intermediate
energies~\cite{VAN06,picap}.

After the absorption of the gauge boson, we follow the path of the
ejected nucleon through its way out of the nucleus using a MC simulation
method to account for the secondary collisions. Details on the MC
simulation can be found in \cite{NVV06}. This MC method  was
designed for single and multiple nucleon and pion emission reactions
induced by pions~\cite{pion1,VicenteVacas:1993bk} and has been
successfully employed to describe inclusive $(\gamma, \pi)$, $(\gamma,
N)$, $(\gamma, NN)$,..., $(\gamma, N\pi)$,...~\cite{CO92bis,COV94},
$(e, e^\prime \pi)$, $(e, e^\prime N)$, $(e, e^\prime NN)$,..., $(e,
e^\prime N\pi)$,...~\cite{GNO97bis} reactions in nuclei or the neutron
and proton spectra from the decay of $\Lambda$
hypernuclei~\cite{Ramos:1996ik}. Thus, we are using a quite robust and
well tested MC simulator.

\section{Results and Concluding Remarks}

\begin{figure}[tbh]
\begin{center}
\includegraphics[scale=0.7, bb=50 350 540 790]{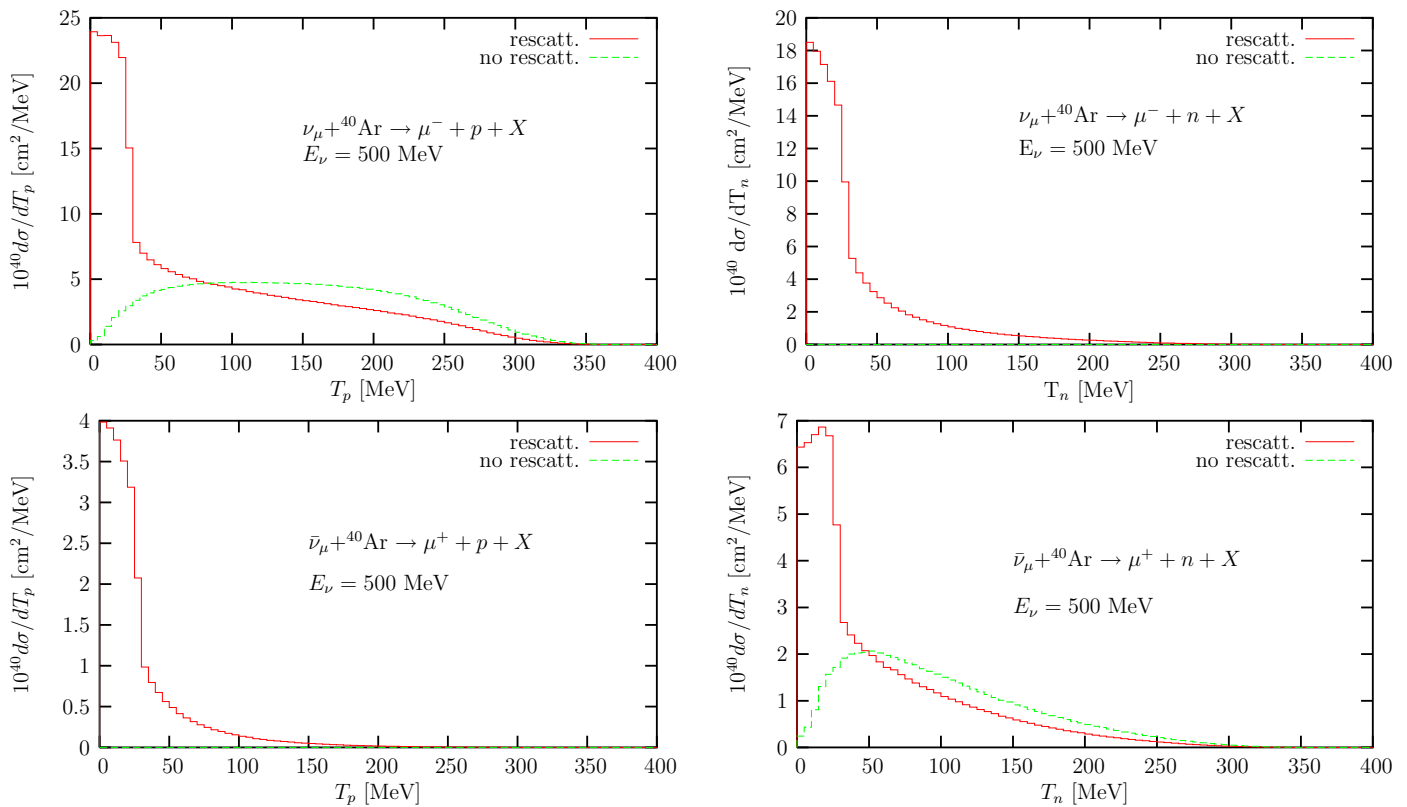}
\caption{ Charged current $^{40}Ar(\nu_{\mu},\mu^-+N)$ (upper panels) and 
$^{40}Ar(\bar{\nu}_{\mu},\mu^++N)$ (lower panels) cross sections as a function
of the kinetic energy of the final nucleon for an incoming neutrino or
antineutrino energy of 500 MeV. Left and right panels correspond to
the emission of protons and neutrons respectively. The dashed histogram 
shows results without nucleon rescattering and the solid one the full model.
}\label{fig:res4}
\end{center}
\end{figure}

The nucleon spectra produced by CC processes induced by muon neutrinos
and antineutrinos of 500 MeV are shown in Fig.~\ref{fig:res4} for
argon. Of course neutrinos only interact via CC with neutrons and
would emit protons, but these primary protons interact strongly with
the medium and collide with other nucleons which
are also ejected. As a consequence there is  a reduction of the flux of high 
energy protons but a large number of secondary nucleons, many of them neutrons,
of lower energies appear. Our cascade model does not 
include the collisions of nucleons with kinetic energies below 30 MeV. Thus,
the results at those low energies are not meaningful and are shown for 
illustrative purposes only in Fig.\ref{fig:res4}.

The flux reduction due to the quasielastic NN interaction can be easily
accommodated in optical potential calculations. However in those calculations
the nucleons that interact are lost when in the physical process they simply
come off the nucleus with a different energy and angle, and may be charge, and 
they must be taken into account.

\begin{figure}[tbh]
\begin{center}
\includegraphics[scale=0.7,  bb= 50 350 540 790]{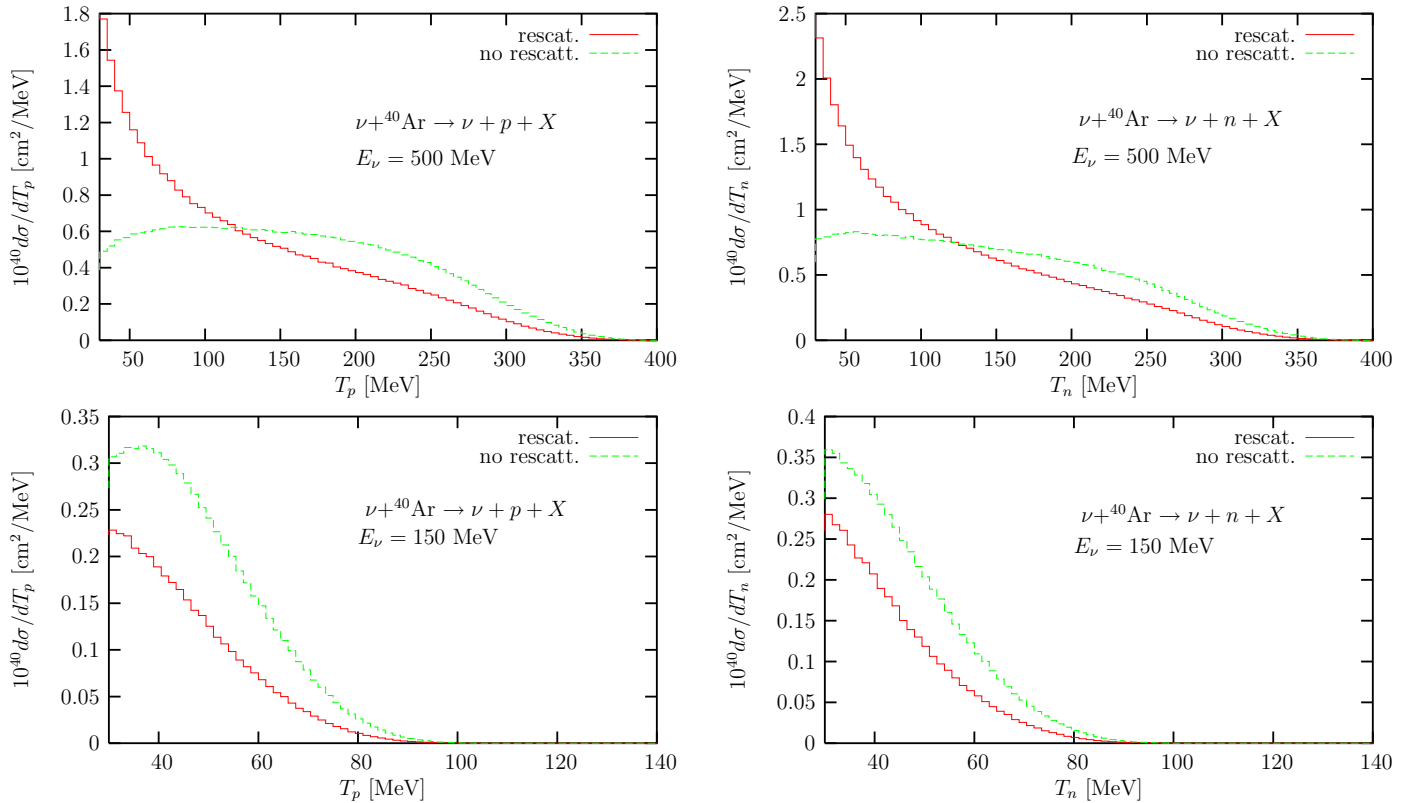}
\caption{ Neutral current
$^{40}Ar(\nu,\nu+N)$ at 500 MeV (upper panels) and 150 MeV (lower
panels) cross sections as a function of the kinetic energy of the
final nucleon. Left and right panels correspond to the emission of
protons and neutrons respectively. The dashed histogram shows results
without rescattering and the solid one the full model.
}\label{fig:res5}
\end{center}
\end{figure}

\begin{figure}[tbh]
\begin{center}
\includegraphics[scale=0.7,  bb= 50 350 540 790]{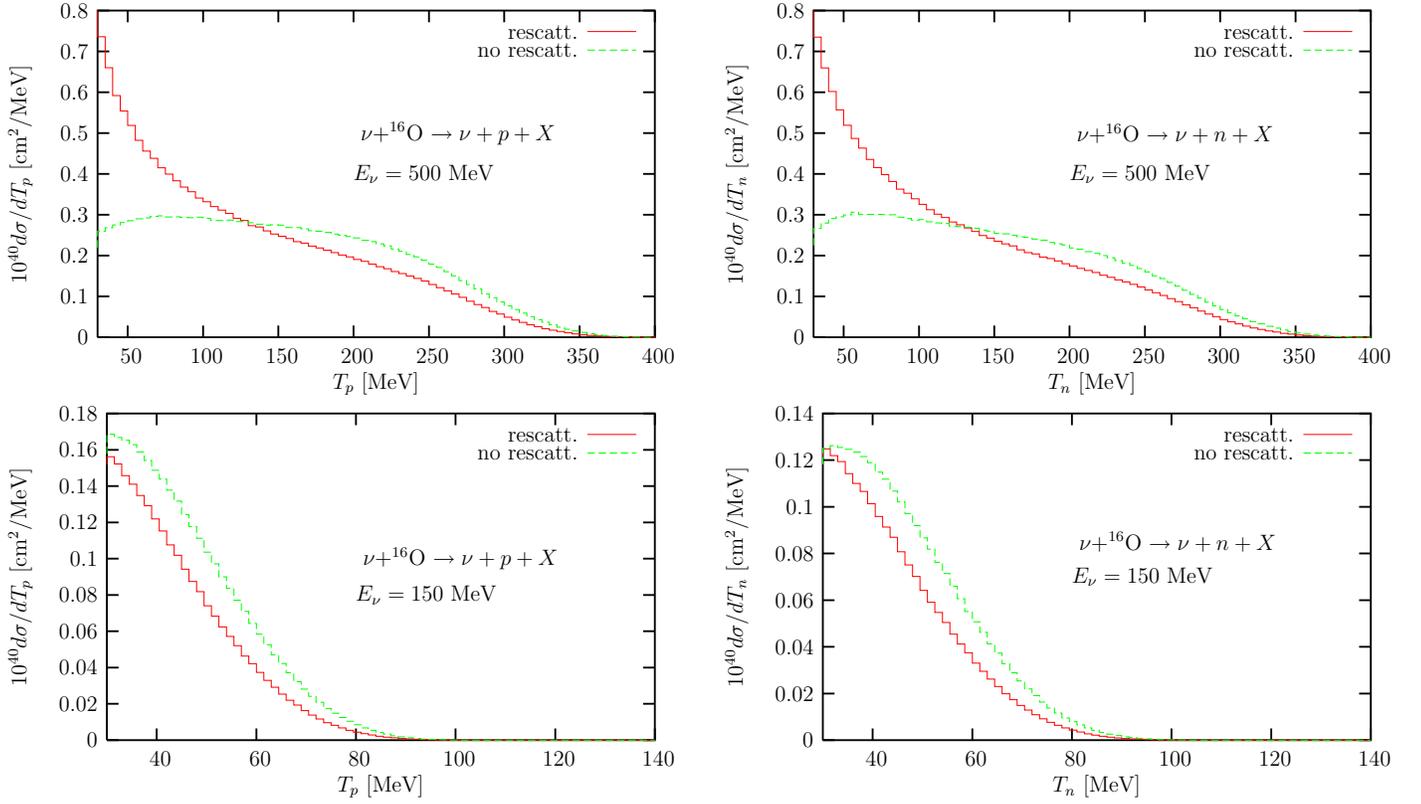}
\caption{  Same as Fig.\ref{fig:res5} for oxygen.}
\label{fig:res5b}
\end{center}
\end{figure}

The energy distributions of nucleons emitted after a NC interaction are shown
in Figs.~\ref{fig:res5} and \ref{fig:res5b}. In Fig.~\ref{fig:res5}, we show 
the results for $^{40}$Ar  at two different energies. In both cases we find the
large effect of the rescattering of the nucleons.
For 500 MeV neutrinos the rescattering of the outgoing nucleon
produces a depletion of the higher energies side of the spectrum, but
the scattered nucleons clearly enhance the low energies region.  For
lower neutrino energies, most of the nucleons coming from nucleon
nucleon collisions would show up at energies below the 30 MeV cut.

As expected, the rescattering effect is smaller in lighter nuclei as
can be seen in Fig.~\ref{fig:res5b} for oxygen. In all cases the final
spectra of protons and neutrons are very similar.

\begin{figure}[tbh]
\begin{center}
\includegraphics[scale=0.7,  bb= 50 560 540 790]{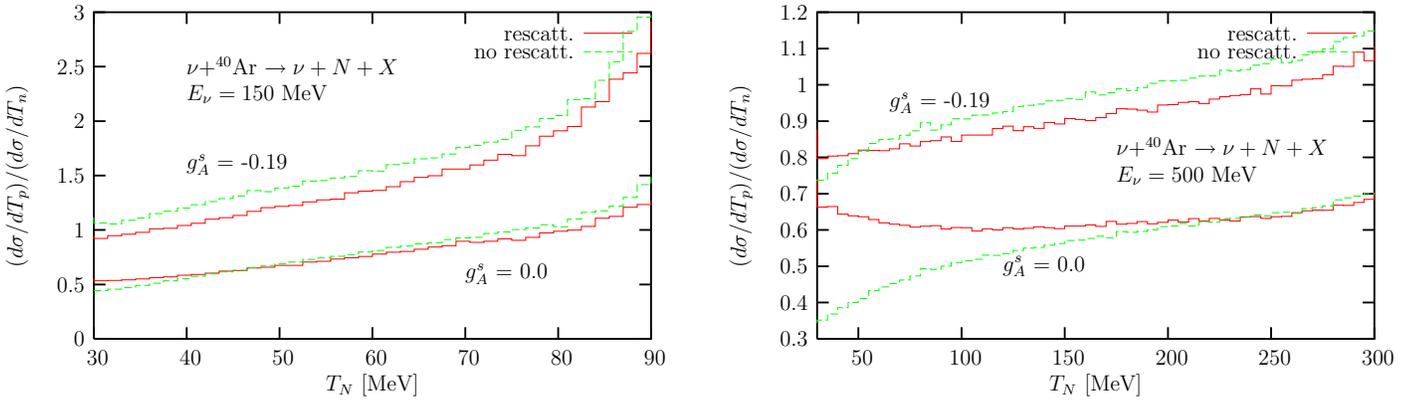}
\caption{ Ratio of $d\sigma/dE$ for protons over that for
neutrons for $E_\nu=150$ MeV and  $E_\nu=500$ MeV in the reaction
$\nu+^{40}Ar\to \nu'+N+X$ as a function of the nucleon kinetic energy.
Dashed histogram: without nucleon rescattering.
Solid histogram: full model.}
\label{fig:res11}
\end{center}
\end{figure}

The ratio of proton to neutron QE cross section could be very
sensitive to the strange quark axial form factor of the nucleon, and
thus to the $g^s_A$ parameter~\cite{Garvey:1992qp,Horowitz:1993rj,Alberico98,
vanderVentel:2003km}. The sensitivity to the collisions of the final
nucleons is larger for both heavier nuclei and for larger energies of
the neutrinos. In Fig.~\ref{fig:res11}, one can  clearly appreciate 
the importance of the secondary nucleons at the low energies side of
the spectrum.

\begin{acknowledgments}
 This work was supported by DGI and FEDER funds, contracts
BFM2005-00810 and BFM2003-00856, by the EU Integrated Infrastructure
Initiative Hadron Physics Project contract RII3-CT-2004-506078 and by
the Junta de Andaluc\'\i a (FQM-225).
\end{acknowledgments}

\end{document}